\documentclass[12pt]{iopart} 
\usepackage[margin=1in]{geometry}                
\usepackage{graphicx}
\usepackage{amssymb}
\usepackage{epstopdf}
\usepackage{color}
\def\artitle#1{} 
\usepackage{Tomspeak} 
\long\def\omitt#1{}

\def\tw#1{}

\begin{document}
\title{Force focusing in confined fibers and sheets}
\author{ Victor Romero$^1$\footnote{
Present address:
Physique et M\'ecanique des Milieux H\'et\'erog\`enes, \'Ecole Sup\'erieure de Physique et de Chimie Industrielles, 10 rue Vauquelin Cedex 5, 75231 Paris, FRANCE}, 
Enrique Cerda$^1$,T. A. Witten$^2$ and Tao Liang$^2$}
\address{$^1$ Departamento de F\'{\i}sica, Universidad de Santiago, Av. Ecuador 3493, Santiago, Chile}
\address{$^2$ James Franck Institute, University of Chicago, Chicago IL 60637 USA.}
\ead{t-witten@uchicago.edu}

\begin{abstract}
A thin fiber or sheet curled into a circular container has a detached region whose shape and force ratios are independent of material properties and container radius. We compute this shape and compare it with experiments.  The discrete forces acting at either end of the detached region have a ratio that depends only on the length of the fiber or sheet relative to the circle radius.  We calculate this force ratio in three regimes of circle radius.
\end{abstract}
\vfill
A sheet of office paper coiled into a mailing tube hugs the wall of the tube in order to minimize its bending.  But the contact with the wall is incomplete;  near the edge, the paper detaches or takes off from the wall and rejoins the cylinder only at the edge.  Such detachment is a commonplace feature of coiled sheets or fibers small and large.  Here we show that the detached region has a universal shape that touches down at an angle of 24.1 degrees.  Moreover, the takeoff point experiences a focused force controlled by the length of the fiber or sheet. 
\hbox to \hsize{ \vbox{\multiply \hsize by 8 \divide\hsize by 17 {\small 
Figure 1 Inset: sheet of office paper coiled in a cardboard tube showing the detachment from the tube at upper right and touchdown at the bottom.  Main figure: A) End view of a long, .02 cm-thick  mica strip (black line overlaid with dashed white line) coiled inside a 4.4-cm-diameter cylinder.   White dashed line shows predicted shape of the detached region.  Subtended angle $\beta$ and touchdown angle $\alpha$ are indicated.  B)  Measured touchdown angle $\alpha$ for different materials and confining radii $R$, illustrating universality of $\alpha$.   $\bullet$: amorphous metal ribbon of thickness .002 cm and  width 0.5 cm; $\blacklozenge$: mica strip of .02 cm thickness and 1 cm width.  Upper error bound indicates the angle at the contact point; lower error bound is the angle extrapolated from the inner surface of the strip to the boundary. Thick horizontal line indicates the predicted universal value of $\alpha$.   C) Takeoff force $T$ relative to touchdown force $P$ vs confining radius $R$, scaled by half-length $S$.}
}\hfill
\vbox{\divide\hsize by 2  \hskip -.2in\includegraphics[width=3.5in]{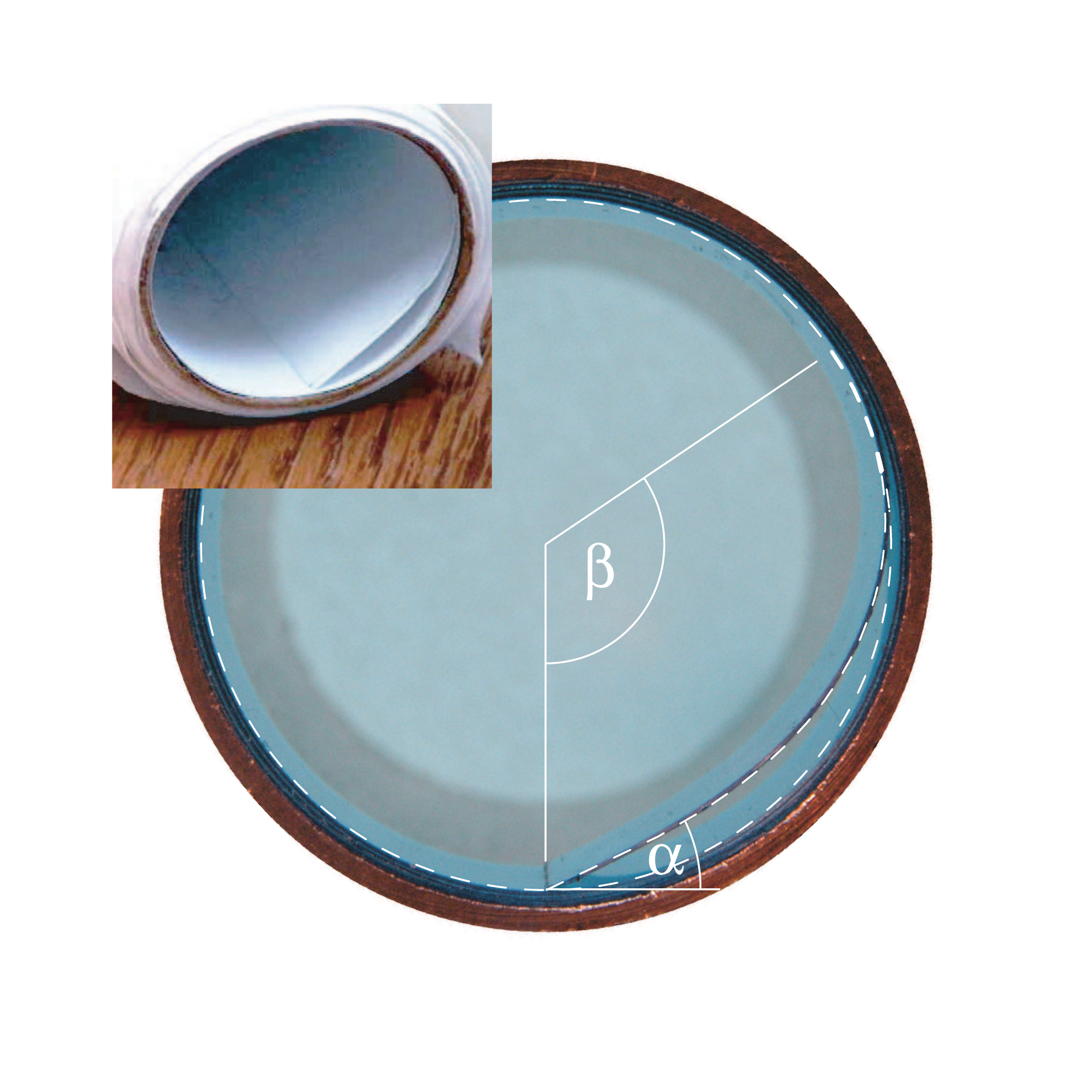}
\vskip -3in \rightline{A)\quad} \vskip 2.3in
\hskip -.2in
 \includegraphics[width=3.5in]{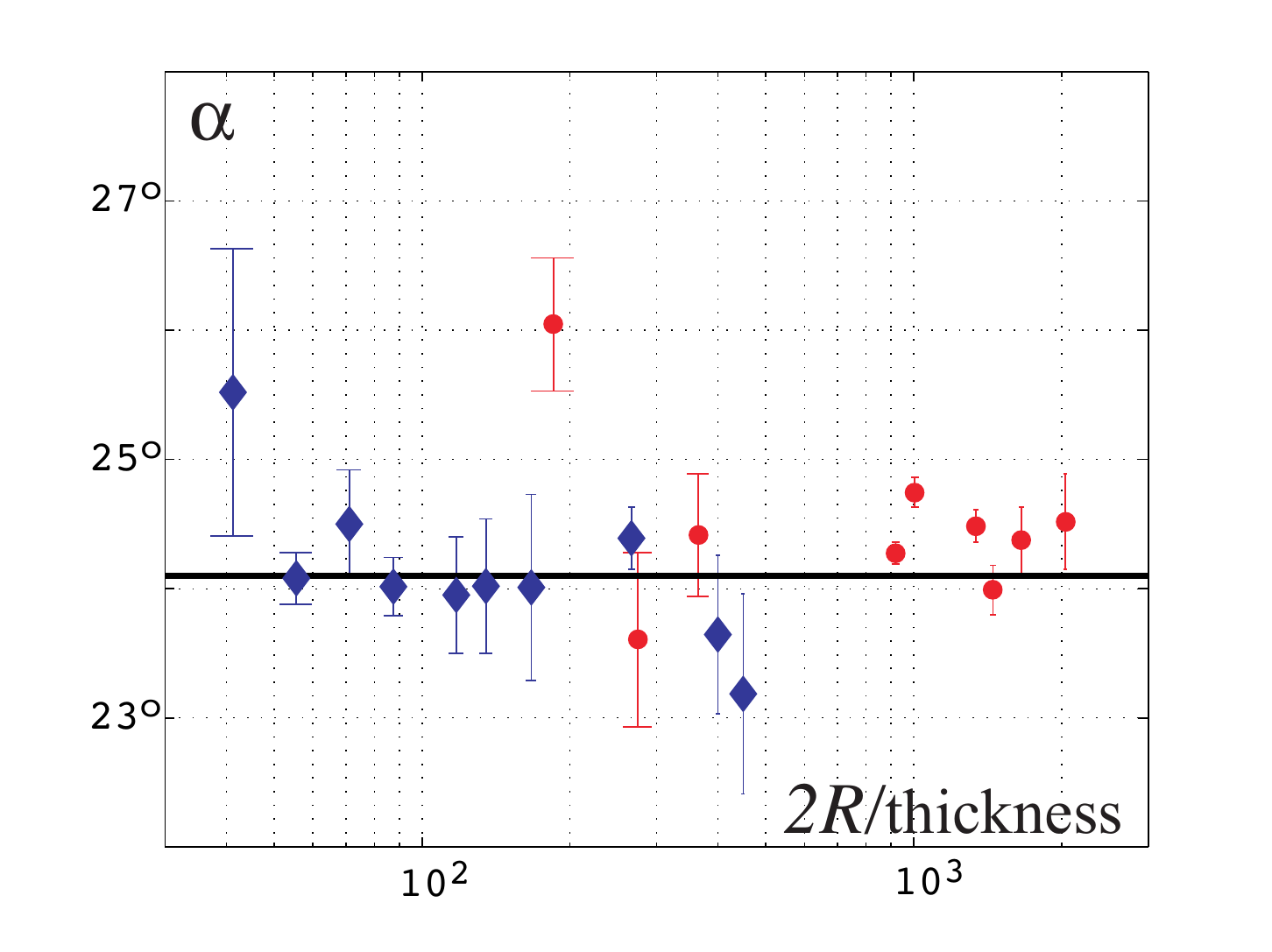}
\vskip -2.4in \rightline{B)\quad} \vskip 2.1in
 \includegraphics[width=3in]{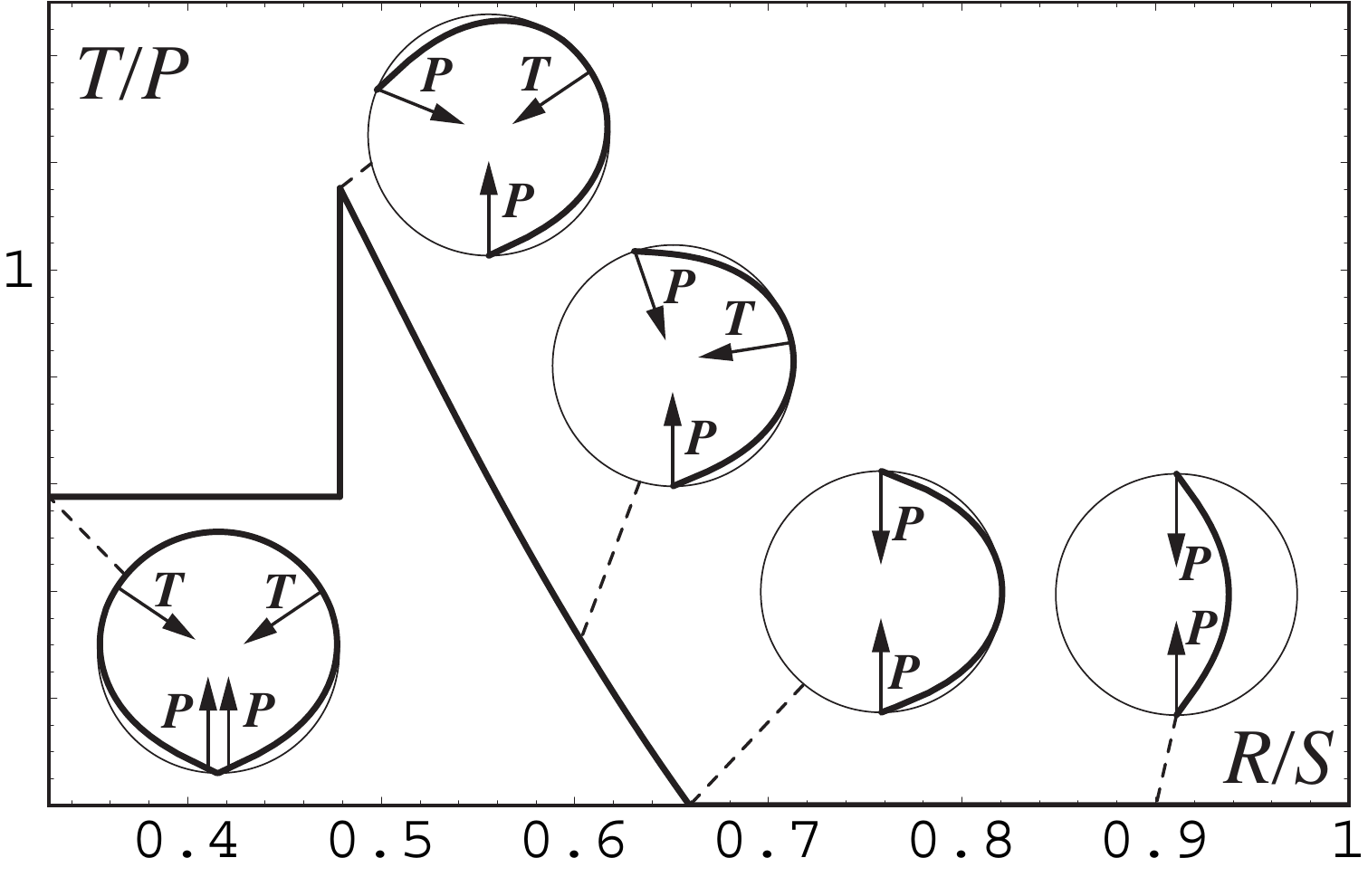}
\vskip -2.0in \rightline{C)\quad} \vskip 1.7in
}}
\bigskip

To explain\cite{Cerda.long} this behaviour, we consider a fibre confined in a circular ring of radius $R$; the same reasoning applies to a sheet confined in a cylinder.  The detached region is necessary because the fibre must be uncurved at the touchdown point.  To bend the fiber at any point requires an external torque around that point\cite{Love.elastica}.  However, the external force $P$ at the touchdown point exerts no torque.  Since the end of the fiber is uncurved, it must angle inward in order to remain in the ring.  Thus a segment of some length $L$ must be detached.  The universal properties reported below arise from the mechanical equilibrium of this detached segment.  

The fiber curves so as to minimize its mechanical energy.   \tw{must revise to include angle $\beta$.  Must take the bottom to be the touchdown point.} We describe the fiber shape by giving its deflection angle $\phi(s)$ from the vertical at a distance $s$ from the takeoff point.  This point $s = 0$ is at angle $\beta$ to the vertical axis, so that $\phi(0) = \beta - \pi/2$.  (Figure 1a).  The end of the fiber, at the bottom of the circle, is at distance $s = L$  from the takeoff point, and $\phi(L) = \alpha - \pi/2$.  The upward touchdown force $P$ acts normal to the circle; thus, frictional forces are not included.  The energy per unit length then consists of the deformation energy, proportional to the square of the curvature $d\phi/ds$, and the work done against $P$. The energy $E$ is thus given by $E =\integral_0^L ds~ [\frac 1 2 B (d\phi/ds)^2 + P\cos(\phi)]$, where $B$ is the bending stiffness of the fiber\cite{Love.elastica}.  The $\phi(s)$ that minimizes such an energy must satisfy the Euler-Lagrange equation\cite{Marion.Thornton}: $Bd^2\phi/ds^2~ +~ P \sin  \phi = 0$.  The fiber at the takeoff point has the same angle and curvature as the adjacent fiber lying against the circle.  This determines $\phi(0) (= \beta - \pi/2)$ and $d \phi/ds~ |_0 (= 1/R)$.  \omitt{See end for crosscheck on signs}
 The value of $P$ must be chosen to make the end of the fiber lie on the constraining circle.   These constraints uniquely determine the fiber shape\cite{Love.elastica}.

The detached region subtends an angle $\beta$ that is independent of  the nature of the fiber or the size of the ring.  That is, it is independent of the bending stiffness $B$ and the confining radius $R$.  Indeed, there is no way to combine $B$, which involves energy, with $R$, which does not, in a way that gives a pure number.  Likewise, the touchdown angle $\alpha$, the relative arc length $L/R$ and the shape of the detached region are independent of $B$ and $R$ (Figure 1B)

The ring exerts a discrete kickoff force denoted $T$ between fiber and ring at the takeoff point.  The tensile counterpart of this force is familiar in de-lamination of multilayer sheets  \cite{delamination}. This kickoff force is needed to balance the touchdown force $P$.  The detached region experiences a tangential pushing force from the remainder of the fiber.  However $P$ also has a component normal to the takeoff surface, and the ring must supply this normal force.  Thus $T = -P \cos \beta $, so that $T/P = 0.5763... $; the fiber functions as a precision force divider.  

Short fibers behave differently from the longer fibers discussed above.   A ring with radius $R$ can accommodate a fiber of half-length $S \le R$ without bending.   If R decreases and becomes smaller than S, the fiber bends while continuing to touch the ring at either side.  Once $R$ has decreased to a value $R_1 = 0.659... S$, the fiber touches the ring at its midpoint.  As $R$ decreases further, the fiber is pressed harder against the ring, its curvature radius at the midpoint $R_c(0)$ increases, the kickoff force $T$ increases as the touchdown points slide away.  Finally $R$ reaches a value $R_2 = 0.478 ... S$ where the radius $R_c$ has increased to $R$.  Now the fiber has begun to lie along the ring.  Further decrease in $R$ simply expands this contacting region so that the two takeoff points move away from the midpoint of the fiber.  The two detached regions then take on the long-fiber shape treated above (Figure 1C)

 It is easy to confine a straight fiber in a circle of known radius.  This geometry could prove useful for making precisely controlled shapes and forces on the macro- and micro scale.  We expect this force-focusing mechanism to be useful where controlled forces need to be exerted on nanometer-scale objects such as surfactant vesicles or biomolecules.  The forcing fiber can be {\it e. g.}  a carbon nanotube \cite{nanotube}, a biofilament \cite{biofilaments}, or a wormlike micelle\cite{wormlike}.  The use of different boundary shapes or deformable boundaries add additional control possibilities.

\tw{don't pule}
\ack
 We thank Juliano Denardin for providing the metal strip used in the experiments, and Sidney Nagel and Aaron Dinner for useful comments.  This work was made possible by the Chicago-Chile Materials Collaboration supported by the National Science foundation under Award Number DMR-0303072.   EC and VR acknowledge the support of Anillo N$^\circ$ ACT 15, FONDAP 11980002, and Fondecyt grant 1050083. TL acknowledges support from the National Science Foundation's MRSEC Program under Award Number DMR-0213745 .
 
\end{document}